\documentclass[a4paper,11pt]{article}
\usepackage{pos}
\usepackage{mathtools}
\usepackage{amsmath}

\title{Model Independent Bounds on Left-Right Gauge Boson Masses from LHC Run 2 and Flavour Observables}
\ShortTitle{Bounds on LR Gauge Boson Masses from LHC and Flavour Observables}

\author*[a]{Sergio Ferrando Solera}
\author[a]{Antonio Pich}
\author[a]{Luiz Vale Silva}

\affiliation[a]{Departament de Física Teòrica, Instituto de Física Corpuscular,
Universitat de València – Consejo Superior de Investigaciones Científicas,\\
  Parc Científic, Catedrático José Beltrán 2, E-46980 Paterna, Valencia, Spain}

\emailAdd{Sergio.Ferrando@ific.uv.es}
\emailAdd{Antonio.Pich@ific.uv.es}
\emailAdd{Luiz.Vale@ific.uv.es}

\abstract{Left-Right Models (LRMs) are one of the most relevant extensions of the Standard Model (SM) of particle physics. They introduce an extended gauge sector and can restore parity (\(\mathcal{P}\)) or charge conjugation (\(\mathcal{C}\)) symmetries at high enough energies. These theories can be embedded in other more fundamental ones with larger gauge groups. Consequently, the restoration of the \(\mathcal{C}\) or \(\mathcal{P}\) symmetries can be pushed towards higher energy scales compared to the scale of the Spontaneous Symmetry Breaking (SSB) of the LRM gauge group. We study three LRMs with different specific realizations of the scalar sector without imposing any additional discrete symmetry on the theory. We present bounds on the masses of the new gauge bosons using data from the LHC Run 2 and study rare meson decays,  discussing the structure of the right-handed quark mixing matrix and the impact of the neutrino and scalar sectors. Collider bounds valid for specific LRM realizations are alleviated bringing New Physics (NP) effects in flavour observables closer to an observable level.}

\FullConference{%
  42nd International Conference on High Energy Physics\\
 18-24 July 2024\\
 Prague Congress Centre, Prague, Czech Republic
}

\begin{document}
\maketitle

\section{Introduction}

The original motivation for the LRMs was to find a natural extension of the gauge group of the SM that allows for the restoration of parity \cite{Original_Ref_1, Original_Ref_2}. For this purpose, a right-handed counterpart of \(\mathrm{SU}(2)_{\mathrm{L}}\) was considered leading to

\begin{equation}
    \label{LR_Group}
    G_{\mathrm{LR}}\coloneqq \underbrace{\mathrm{SU}(3)_{\mathrm{QCD}}}_{g_S}\otimes\underbrace{\mathrm{SU}(2)_{\mathrm{L}}}_{g_L}\otimes\underbrace{\mathrm{SU}(2)_{\mathrm{R}}}_{g_R}\otimes\underbrace{\mathrm{U}(1)_{X}}_{g_X},
\end{equation}

\noindent where the \(g\)s are the gauge couplings of the theory and the generator \(X\) is given by \(X=\left(B-L\right)/2\), with \(B\) and \(L\) representing the baryon and lepton numbers, respectively.  We will assume that this symmetry is spontaneously broken into \(G_{\rm SM}\coloneqq\mathrm{SU}(3)_{\mathrm{QCD}}\otimes\mathrm{SU}(2)_{\mathrm{L}} \otimes \mathrm{U}(1)_{Y}\) at the LR energy scale  \(v_R\) which, for phenomenological reasons, is supposed to be much higher than the electroweak scale (\(v_R\gg v_{\rm EW}\)). Analogous to the SSB of the SM, this will introduce a mixing angle \(\gamma\) between the neutral gauge fields \(W_R^3\) and \(W_X\) associated with the groups \(\mathrm{SU}(2)_{\mathrm{R}}\) and \(\mathrm{U}(1)_{X}\). As a second step, \(G_{\rm SM}\) will be spontaneously broken into \(\mathrm{U}(1)_{\rm QED}\) at the energy scale \(v_{\rm EW}\) with the appearance of the usual Weinberg angle \(\theta_W\), which induces the mixing between the states \(W_L^3\) and \(B\):

\begin{equation}
    \label{SSB_Mechanism}
    G_{\mathrm{LR}}\xrightarrow{v_R,\,\gamma\left(W_R^3,W_X\right)}G_{\rm SM}\xrightarrow{v_{\rm EW},\,\theta_W\left(W_L^3,B\right)}\mathrm{U}(1)_{\rm QED}.
\end{equation}

\noindent The coupling constants and the mixing angles are related by the expression

\begin{equation}
    \label{Coupling_Constants}
    e=g_L\sin\theta_W=g_R\sin\gamma\cos\theta_W=g_X\cos\gamma\cos\theta_W,
\end{equation}

\noindent where \(e\) is the coupling constant associated with \(\mathrm{U}(1)_{\rm QED}\). To restore the parity symmetry we need to impose the condition \(g_L=g_R\), which leads to the constraint \(\sin\gamma=\tan\theta_W\), satisfied for \(\gamma\approx33^{\circ}\). Nevertheless, it is well-known that the gauge group \(G_{\rm LR}\) can be embedded in other Beyond the Standard Model (BSM) groups, like \(\mathrm{SO}(10)\) \cite{Embedding}, meaning that the restoration of parity could be pushed towards higher energy scales and the values of \(g_L\) and \(g_R\) would be determined by the running of the coupling constants of those more fundamental theories at lower energies. Therefore, we will study the case where no additional discrete symmetries are imposed on the LRMs. Even in this context we can constrain the values of the mixing angle \(\gamma\) if we assume that we are in the perturbative regime. For instance, if we impose that \(g_R,g_X\lesssim1\) we get \(20^{\circ}\lesssim\gamma\lesssim70^{\circ}\).

The fermion content of the LRMs is usually given by the left- and right-handed doublets of quarks and fermions: \(\left(U_L,D_L\right)^{\rm T}\sim\left(\textbf{3},\textbf{2},\textbf{1}\right)_{1/6}\), \(\left(U_R,D_R\right)^{\rm T}\sim\left(\textbf{3},\textbf{1},\textbf{2}\right)_{1/6}\), \(\left(e_L,\nu_L\right)^{\rm T}\sim\left(\textbf{1},\textbf{2},\textbf{1}\right)_{-1/2}\) and  \(\left(e_R,\nu_R\right)^{\rm T}\sim\left(\textbf{1},\textbf{1},\textbf{2}\right)_{-1/2}\). It is important to remark that the addition of the right-handed neutrino is necessary and its nature as a Majorana or a Dirac particle will depend on the concrete SSB mechanism. Apart from the \(V_L^{\rm CKM}\) quark mixing matrix of the SM, now we find its right-handed counterpart \(V_R^{\rm CKM}\) and the analogous lepton mixing matrices \(V_L^{\rm PMNS}\) and \(V_R^{\rm PMNS}\). 
Furthermore, if we do not impose any additional discrete symmetry on the theory, the right-handed mixing matrices will be completely general unitary matrices unrelated to the left-handed ones \cite{General_VR}. 

We have considered three different SSB mechanisms, including the two most studied ones, and a third one which was introduced in \cite{Our_Paper}:

\begin{enumerate}
    \item \textit{Two Doublets and One Bidoublet:} here the scalar sector consists of two doublets \(\chi_L\sim\left(\textbf{1},\textbf{2},\textbf{1}\right)_{1/2}\) and \(\chi_R\sim\left(\textbf{1},\textbf{1},\textbf{2}\right)_{1/2}\), and a bidoublet \(\Phi\sim\left(\textbf{1},\textbf{2},\textbf{2}\right)_{0}\). In this model all of the fermions are Dirac particles, so right-handed neutrinos are extremely light.
    \item \textit{Two Triplets and One Bidoublet:} in this model the scalar sector consists of two triplets 
    \(\Delta_L\sim\left(\textbf{1},\textbf{3},\textbf{1}\right)_{1}\) and \(\Delta_R\sim\left(\textbf{1},\textbf{1},\textbf{3}\right)_{1}\), and the bidoublet \(\Phi\). Here neutrinos are Majorana fermions and the heavy ones have a mass proportional to the LR scale \(v_R\), while the mass of the light ones is proportional to \(m_{\nu_L}\propto v_{\rm EW}^2/v_R\), implementing a Seesaw Mechanism. 
    \item \textit{Two Doublets Effective Model:} if we only consider the two doublets \(\chi_L\) and \(\chi_R\) the masses of the fermions need to be generated by dimension-five operators and it is possible to write a Majorana mass term for neutrinos. Considering effective interactions is very natural in this context since we are assuming that these models are embedded in other more fundamental theories with a new energy scale \(\Lambda_{\rm NP}\). The masses of the charged fermions are proportional to \(m_{\rm cf}\propto v_{\rm EW}v_R/\Lambda_{\rm NP}\), for the heavy neutrinos we have \(m_{\nu_h}\propto v_R^2/\Lambda_{\rm NP}\) and for the light ones \(m_{\nu_l}\propto v_{\rm EW}^2/\Lambda_{\rm NP}\). 
\end{enumerate}

\section{Flavour Observables}

The main goal of our work is to obtain bounds on the masses of the new gauge bosons and, consequently, on the LR scale without imposing any additional discrete symmetry on the theory and independently of the values of the gauge couplings, the masses of the heavy neutrinos (if present) and the masses and interactions of the scalars (although it may depend on the particular scalar content). We will only assume that the theory is perturbative and that \(v_R\gg v_{\rm EW}\).

For this purpose, we might try to use data coming from flavour observables, but we will later show that, as a first step, we can find simpler alternatives. For instance, it is well-known that  the theoretical predictions of these processes often require the calculation of box diagrams, whose amplitude is proportional to the masses of the internal fermions due to the GIM mechanism. In the effective model this amplitude is proportional to \(\Lambda_{\rm NP}^{-2}\), so the complete calculation requires the addition of all possible dimension 6 operators thus requiring more details about the UV completion. Furthermore, in processes like \(b\to s\gamma\) where only the penguin diagram is present, the leading NP contribution coming from the gauge fields is proportional to the mixing between the \(W_L\) and the \(W_R\). Since there is no mixing in the effective model \cite{Our_Paper}, the leading contribution will be proportional to \(M_{W_R}^{-4}\) making it completely negligible.

For the other two scenarios, we can illustrate the situation considering the processes \(b\to s\gamma\) and \(b\to sll\). In these theories, if we do not impose any additional symmetry, the Higgs potential is highly non-trivial since it contains a huge number of free parameters. Nevertheless, given the bounds on the masses of the neutral scalars coming from Flavour Changing Neutral Currents (FCNCs), we can make the assumption that they are much heavier than the gauge bosons and neglect their contribution as a first approximation. For the sake of illustration, we will only consider the model with two doublets and the bidoublet, where right-handed neutrinos are light. Thus, following \cite{Flavour_Amplitudes}, we have computed the new physics contribution to the Wilson coefficients

\begin{equation}
    \label{LR_C7}
    \left.C_7\right|_{\mathrm{NP}}=\frac{1}{\sqrt{2}}4em_bG_F\sum_{i=u,c,t}\frac{g_R}{g_L}\frac{m_i}{m_b}\sin\xi_W e^{-i\lambda}\left(V_L^{\mathrm{CKM}}\right)_{is}^{\ast}\left(V_R^{\mathrm{CKM}}\right)_{ib}\tilde{F}(x^{f_i}_{W_L}),
\end{equation}

\begin{equation}
    \label{LR_C7_prime}
    \left.C_7^{\prime}\right|_{\mathrm{NP}}=\frac{1}{\sqrt{2}}4em_bG_F\sum_{i=u,c,t}\frac{g_R}{g_L}\frac{m_i}{m_b}\sin\xi_W e^{i\lambda}\left(V_R^{\mathrm{CKM}}\right)_{is}^{\ast}\left(V_L^{\mathrm{CKM}}\right)_{ib}\tilde{F}(x^{f_i}_{W_L}),
\end{equation}

\begin{equation}
    \label{C9prime}
     C_9^{\prime l}= \frac{4}{9}\left(\frac{eg_R}{M_{W_R}}\right)^2\sum_{i=u,c,t}\left(V_R^{\mathrm{CKM}}\right)^{\ast}_{is}\left(V_R^{\mathrm{CKM}}\right)_{ib}\ln\left(\frac{m_i}{m_0}\right),
\end{equation}

\noindent with \(\tilde{F}\left(x\right)\coloneqq\left(-20+31x-5x^2\right)/12\left(x-1\right)^2-x\left(2-3x\right)\ln x/2\left(x-1\right)^3\), \(x_{W_L}^{f_i}\coloneqq m_{f_i}^2/M^2_{W_L}\), \(m_0\) the mass of a particular up-type quark, \(G_F\) the Fermi constant and \(\xi_W\) and \(\lambda\) the parameters of the \(W_L-W_R\) mixing \cite{Our_Paper}. The normalization that we are taking is 

\begin{equation}
    \label{Dipole_Lagrangian}
    \mathcal{L}_{\mathrm{Eff}}=\frac{1}{16\pi^2}\left\{C_7\left(\bar{s}\sigma^{\mu\nu}P_R b\right)F_{\mu\nu}+C_7^{\prime}\left(\bar{s}\sigma^{\mu\nu}P_L b\right)F_{\mu\nu}+C_9^{\prime l}\left(\bar{s}\gamma_{\mu}P_Rb\right)\left(\bar{l}\gamma^{\mu}l\right)\right\}.
\end{equation}

All of the other Wilson coefficients are of the order of \(v_R^{-4}\) so we will neglect them. As we can see, even with all of the assumptions that we have made, the contributions to the flavour observables still depend on many free parameters due to the appearance of \(V_R^{\rm CKM}\). Thus, obtaining model independent bounds from these processes will require a global fit combining multiple flavour observables. In order to know what we could expect, we can impose \(g_L=g_R\), consider a diagonal structure for \(V_R^{\rm CKM}\) and take \(\lambda=0\) and \(\sin\xi_W=\left(v_{\rm EW}/v_R\right)^2\). In this case, from the bounds on \(\left.C_7\right|_{\mathrm{NP}}\) obtained in \cite{Bounds_C7}, we get \(v_R\sim 7\,{\rm TeV}\). We will see that this is similar, but not as strong as the bounds that we have obtained from collider physics in the scenarios that have light neutrinos (case \(\Phi+\chi_{L,R}\) it Tab. \ref{Summary_Tab}).

\section{Collider Bounds}

\begin{figure}
    \centering
    \includegraphics[scale=0.31]{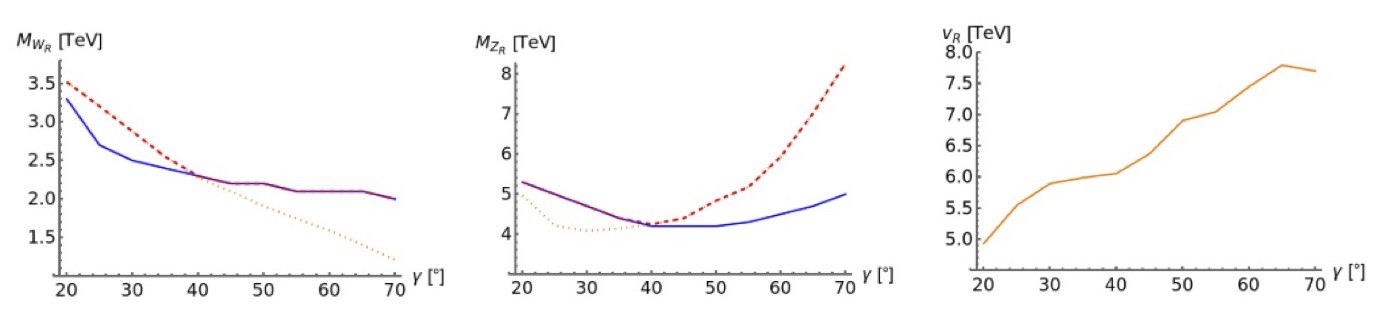} 
    \caption{Bounds on \(M_{W_R}\) (left panel), \(M_{Z_R}\) (center panel) and \(v_R\) (right panel) for different values of the mixing angle \(\gamma\) in the triplet model. In the plots of the masses of the gauge bosons the solid blue line represents the direct bound, the dashed yellow line the indirect bound and the dotted red line the combined bound.}
    \label{Bounds_Fig}
\end{figure}

We focus our attention on the Drell-Yan production in the LHC of the \(W_R\) and \(Z_R\) gauge bosons. For the \(Z_R\) we study the decay channel \(Z_R\to l^+l^-\) while for the \(W_R\) we consider its decay into two jets or into a charged lepton and a light right-handed neutrino in the model with two doublets and a bidoublet.

We have used the Narrow Width Approximation (NWA), where the cross section can be factorized as \(\sigma\left(pp\to W_R/Z_RX\to f_1\bar{f}_2X\right)\approx \sigma\left(pp\to W_R/Z_RX\right){\rm Br}\left(W_R/Z_R\to f_1\bar{f}_2\right)\), where \(X\) is a state resulting from the strong dynamics in the collision process. For the \(W_R\) we have


\begin{equation}
    \label{Production_WR}
    \sigma\left(pp\rightarrow W_R X \right)\approx\frac{\pi}{6s}g_R\sum_{ij}|\!\left(V_R^{\mathrm{CKM}}\right)_{ij}\!|^2\,\omega_{ij}\!\left(M^2_{W_R}/s,\,M_{W_R}\right),
\end{equation}

\noindent where \(\omega_{ij}\) is a function of the parton distribution function and \(s\) the center of mass energy. As we can see, we have a dependence on \(g_R\) and on the right-handed quark mixing matrix \(V_R^{\mathrm{CKM}}\), in contrast, for the \(Z_R\) there is only a dependence on the mixing angle \(\gamma\) \cite{Our_Paper}. We look for the least constraining lower bound on the masses of the new gauge fields, in the sense that any specific realistic choice of the other parameters of the theory will produce a bigger lower bound. Thus, we can find the structure for the right-handed mixing matrix that leads to the smallest production cross section and, as it is shown in \cite{Our_Paper}, it corresponds to the antidiagonal case. 
In \cite{Our_Paper} it is also shown how to get an upper bound on the total width of the gauge bosons for fixed gauge couplings, which leads to the smallest total cross section of these processes. This is something non-trivial considering the fact that we do not know the masses of the scalars, but the upper bounds only depend on the particular values of the \(\gamma\) angle and on the scalar content of the theory. In the perturbative regime the values of \(\Gamma_V/M_V\) are all below \(10\%\) so the NWA is justified.


In order to get the bounds on the masses we have compared the theoretical cross sections with the data from the LHC Run 2. We have used \cite{CMS_Z_Leptons} for \(Z_R\to l^+l^-\), \cite{ATLAS_W_Jets} for \(W_R\to j_1j_2\) and \cite{ATLAS_W_lnu} for \(W_R\to l\nu_R\). Then, we can get the least constraining lower bounds on the masses of the heavy gauge bosons for different values of the angle \(\gamma\). Nonetheless, these bounds are complementary, because the masses of the \(W_R\) and the \(Z_R\) are related by \(M_{W_R} = M_{Z_R}\cos \gamma=ev_R/(2\cos\theta_W\sin\gamma)\), in the models with doublets, and \(M_{W_R} = M_{Z_R}\cos \gamma/\sqrt{2}=ev_R/(\sqrt{2}\cos\theta_W\sin\gamma)\), in the models with triplets. Thus, a direct bound on one mass leads to an indirect bound on the other (see Fig. \ref{Bounds_Fig}), so we need to combine them. Moreover, from the previous relations we also get constraints on \(v_R\) for different values of \(\gamma\). Finally, taking the value of the mixing angle for which we get the smallest bounds, we obtain the least constraining lower bounds, which are summarized in Tab. \ref{Summary_Tab}.

\begin{table}
    \centering
    \begin{tabular}{|c|c|c|c|}
    \hline
         & \(\Phi+\chi_{L,R}\) & \(\Phi+\Delta_{L,R}\) & \(\chi_L+\chi_R\) \\ \hline
      \(M_{W_R}\left[{\rm TeV}\right]\)   & 4.3 & 2.0 & 2.1 \\ \hline
      \(M_{Z_R}\left[{\rm TeV}\right]\)   & 5.4 & 4.2 & 4.3 \\ \hline
      \(v_{R}\left[{\rm TeV}\right]\)   & 10 & 4.9 & 10 \\ 
    \hline
    \end{tabular}
    \caption{Summary of the bounds on the masses of the heavy gauge bosons and on the LR scale for the three different SSB mechanisms.}
    \label{Summary_Tab}
\end{table}

\section{Conclusions}

We have studied three LR theories (including an effective model) differentiated by their scalar content. The novelty is that we have not imposed any restriction (apart from perturbativity) on the free parameters, like the elements of the right-handed quark mixing matrix \(V_R^{\rm CKM}\) or the coupling constant \(g_R\). We have seen that, in this scenario, obtaining bounds on the masses of the heavy gauge bosons using flavour observables would require a global fit and the expected results may not improve the ones from collider physics when \(V_R^{\rm CKM}\) is diagonal. It is important to stress that the collider bounds on \(M_{W_R}\) and \(M_{Z_R}\) are complementary. This work attests to the high sensitivity that the LHC has achieved.

\section*{Acknowledgments}
We thank Prasanna K. Dhani, Andreas Hinzmann, Greg Landsberg, Jeongeun Lee, Emanuela
Musumeci, Tamara Vázquez Schroeder, and José Zurita for engaging in discussions and providing
their comments. This work has been supported by MCIN/AEI/10.13039/501100011033, grant
PID2020-114473GB-I00; by Generalitat Valenciana, grant PROMETEO/2021/071 and by Ministerio de Universidades (Gobierno de España), grant FPU20/04279. This project has received
funding from the European Union’s Horizon 2020 research and innovation programme under the
Marie Sklodowska-Curie grant agreement No 101031558. LVS is grateful for the hospitality of
the CERN-TH group where part of this research was executed.

\end{document}